\documentclass[twocolumn,showpacs,preprintnumbers,amsmath,amssymb,superscriptaddress]{revtex4}
\usepackage{epsfig}
\usepackage{epstopdf}
\usepackage{graphicx}
\usepackage{dcolumn}
\usepackage{bm}
\usepackage{amsmath}

\newcommand{\atm}{atmospheric}
\newcommand{\nn}{neutrino}
\newcommand{\nns}{neutrinos}
\newcommand{\be}{\begin{equation}}
\newcommand{\ee}{\end{equation}}
\newcommand{\beq}{\begin{eqnarray}}
\newcommand{\eeq}{\end{eqnarray}}

\def\epee{\mathrel{{\epsilon_{ee}}}}
\def\epem{\mathrel{{\epsilon_{e\mu}}}}
\def\epmm{\mathrel{{\epsilon_{\mu\mu}}}}
\def\epet{\mathrel{{\epsilon_{e\tau}}}}
\def\epmt{\mathrel{{\epsilon_{\mu\tau}}}}
\def\eptt{\mathrel{{\epsilon_{\tau\tau}}}}

\def\nue{\mathrel{{\nu_e}}}
\def\numu{\mathrel{{\nu_\mu}}}
\def\nutau{\mathrel{{\nu_\tau}}}

\def \lta {\mathrel{\vcenter{\hbox{$<$}\nointerlineskip\hbox{$\sim$}}}}
\def \gta {\mathrel{\vcenter{\hbox{$>$}\nointerlineskip\hbox{$\sim$}}}}

\def\t13{\mathrel{{\theta_{13}}}}
\def\y12{\mathrel{{\tan^2 \theta_{12}}}}
\def\dm{\mathrel{{\Delta m^2}}}
\def\unitdm{\mathrel{{\cdot 10^{-3}~{\rm eV^2}}}}

\begin{document}

\preprint{LA-UR-06-3973}

\title{Two modes of searching for new neutrino interactions at MINOS}
\author{Alexander Friedland}
 \email{friedland@lanl.gov}
\affiliation{%
Theoretical Division, T-8, MS B285, Los Alamos National Laboratory, Los Alamos, NM 87545
}%

\author{Cecilia Lunardini}
 \email{lunardi@phys.washington.edu}
\affiliation{Institute for Nuclear Theory and University of Washington, Seattle, WA 98195
}%


\begin{abstract}
The SuperKamiokande atmospheric neutrino measurements leave
substantial room for nonstandard interactions (NSI) of \nns\ with
matter in the $\nu_e- \nu_\tau$ sector. Large values of the NSI
couplings are accommodated if the vacuum oscillation parameters
are changed from their standard values. Short and medium baseline
neutrino beams can break this degeneracy by measuring the true
vacuum oscillation parameters with the $\numu$ disappearance mode,
for which the matter effects are negligible or subdominant. These
experiments can also search for the $\nu_e- \nu_\tau$ flavor
changing effects directly, by looking for $\numu -\nue$ conversion
caused by the intervening matter.  We discuss both of these
methods for the case of MINOS.  We find that, while the present
MINOS data on $\numu$ disappearance induce only minor changes on
the constraints on the NSI parameters, the situation will improve
markedly with the planned increase of the statistics by an order
of magnitude. In that case, the precision will be enough to
distinguish certain presently allowed NSI scenarios from the
no-NSI case. NSI per quark of about 10\% the size of the standard
weak interaction could give a $\numu - \nue$ conversion
probability of the order $\sim 10^{-2}$, measurable by MINOS in
the same high statistics scenario. In this $\numu - \nue$ channel,
the small effects of NSI could be comparable or larger than the
vacuum contribution of the small angle $\theta_{13}$. The expected
$\theta_{13}$ bound at MINOS should be more properly interpreted
as a bound in the $\theta_{13}$ -- NSI parameter space.

%
\end{abstract}

\maketitle

\section{Introduction}

Neutrinos offer certain unique advantages for the search of
new forces beyond the Standard Model. These advantages stem from
the fact that  neutrinos are sensitive only to the weakest
known forces, the weak interaction and gravity, and therefore
offer a ``cleaner'' ground -- with respect to other particles --
to look for the subdominant effects of the new, nonstandard
interactions (NSI). The abundance of neutrinos from natural
sources (chiefly the Sun and the interaction of cosmic rays in the
Earth's atmosphere) adds further motivation.

The very same reason that makes them an interesting probe, the
weakness of their interactions, also made the neutrinos
historically difficult to study experimentally. The result is that
to this day some of the neutrino interactions are very poorly
known, with  uncertainties comparable to, or even exceeding, the
standard model predictions. In the case of neutrino-matter
interactions the measurements of neutrino cross sections at CHARM
and NuTeV \cite{Vilain:1994qy,Zeller:2001hh} constrain chiefly the
coupling of the muon neutrino, while leaving the $\nue-\nutau$
channels largely unconstrained (see
\cite{Berezhiani:2001rs,Davidson:2003ha} for a comprehensive
discussion).

Measurements of neutrino matter effects at neutrino oscillation
experiments offer increased sensitivity to neutral current NSI in
the $e - \tau$ sector, down to tens of per cent of the weak
interaction strength
\cite{Fornengo:2001pm,Guzzo:2001mi,Gonzalez-Garcia:2004wg}. Still,
however, certain directions in the parameter space exist where
large NSI are allowed, because their effect on the oscillations of
solar and atmospheric neutrinos is degenerate with that of the
mixings and mass square differences
\cite{Friedland:2004pp,Friedland:2004ah,Friedland:2005vy}.

Logically, the next step towards stronger constraints -- or
discovery -- of neutrino NSI is to break these degeneracies. One
way to do so is to perform precision measurements of the neutrino
mixings and mass squared differences in conditions where matter
effects are negligible or subdominant. Such conditions of ``quasi-vacuum'' are
realized in the $\numu -\nutau$ oscillation channel of the $\numu$
beams used by K2K \cite{Ahn:2002up,Aliu:2004sq} and MINOS
\cite{Adamson:2005qc,talkpetyt}. There, the large $\numu -\nutau$
(vacuum) mixing, $\theta_{23}$, induces oscillations effects of
order unity.  These dominate over the matter effects, which are at
most of order $\sim 10^{-1}$, due to the small matter width along
the beam trajectory.

In a recent paper \cite{Friedland:2005vy} we showed  how the data
from K2K, in combination with the SuperKamiokande atmospheric
neutrino measurements, restrict the allowed region of the NSI
couplings. The conclusion was that,  while improving with respect
to the \atm\ data only, the combined data still allow couplings of
order unity (relative to the Fermi constant, $G_F$).

Here we discuss how the situation changes in light of the recently
reported search for $\numu$ disappearance at MINOS
\cite{talkpetyt}. We also illustrate the increased potential that
MINOS will have with higher statistics. We get that with the
present MINOS data the presence of NSI is neither revealed nor
excluded. The allowed region of the NSI parameters is changed only
slightly.  Qualitative improvements are expected, however, with
the planned increase of statistics by an order of magnitude. The
precision in that case would be such that in certain NSI scenarios
(with the NSI couplings on both $u$ and $d$ quarks of the order of
10\% of the weak interaction), the point of zero NSI would be
allowed only with confidence level higher than 99\%.
%

A second way to search for the flavor-changing $\nue-\nutau$ NSI
is to look for the $\numu - \nue$ conversion \cite{Gago:2001xg,Huber:2001de,Campanelli:2002cc,Blennow:2005qj,Huber:2002bi}. In this channel, the
vacuum contribution to the oscillations of beam neutrinos is due
to the small mixing angle $\theta_{13}$. Considering that $\sin^2
2\theta_{13}\lta 0.14$ \cite{Apollonio:1999ae,Apollonio:2002gd} we see that at K2K and MINOS the
NSI effects could be of the same order or larger than the vacuum
ones, and therefore they could be visible in spite of their
smallness on absolute scale.
In this paper we illustrate this point for the specific case of MINOS,
and show that a conversion probability of $\sim 0.05$, within the
reach of MINOS, could be due to NSI within the currently allowed
region, with couplings per quark of the order of 10\% of the weak
interaction.

Finally, although not discussed here, another important method of
searching for new tau neutrino interactions at MINOS is by
measuring the neutral current event rate in the far detector,
where a large fraction of events should be due to tau neutrinos
generated by oscillations. The combinations of the NSI parameters
probed in this method do not uniquely map to those probed by the
matter effects studied here \cite{Friedland:2005vy}, hence
providing a complementary probe of NSI.

In Sec.~\ref{generalities} we describe the parametrization of the
nonstandard interactions being probed, list various constraints on
these interactions and give the resulting oscillation Hamiltonian.
We also  review the main results of
our earlier study of NSI effects on atmospheric neutrinos  (in Sect.~\ref{sect:atmreview}). In
Sect.~\ref{oscillations} we study the combined NSI sensitivity of
the $\numu\rightarrow\nutau$ mode at MINOS and atmospheric and K2K
data. Sec.~\ref{numunue} is devoted to the $\numu \rightarrow\nue$
appearance channel at MINOS. Finally, Sec.~\ref{summary}
summarizes our findings.


\section{Neutrino oscillations and NSI in the $\nue - \nutau$ sector }
\label{generalities}

\subsection{Effective Lagrangian, bounds, oscillation Hamiltonian}

The neutrino oscillation Hamiltonian in vacuum is dominated by the
atmospheric mass splitting. In the flavor basis $\nue, \numu,
\nutau$, this Hamiltonian is given by
\begin{equation}
  \label{eq:vacHprime}
    H_{\rm vac}\simeq\Delta \left(\begin{array}{ccc}
   - 1 & 0 & 0 \\
   0 & -\cos 2\theta & \sin 2\theta \\
  0  & \sin 2\theta & \cos 2\theta \\
\end{array}\right),
\end{equation}
where $\Delta\equiv \Delta m^2/(4 E_\nu)$, $E_\nu$ is the neutrino
energy and $\Delta m^2$ is the largest mass square difference in
the neutrino spectrum (often called $\Delta m^2_{32}$).
Corrections to $H_{vac}$ due to the smaller, ``solar'' mass
splitting and the remaining two mixing angles produce only
subdominant  effects on the $\numu \rightarrow \nutau$
oscillations (see sec. \ref{oscillations}) and so will be omitted for the moment. See,
{\it e.g.}, \cite{Friedland:2005vy} for the full Hamiltonian
including all these terms.

In matter, the neutrino propagation is affected by refraction
(coherent forward scattering). We parameterize the corresponding
part of the Hamiltonian as:
\begin{equation}
H_{\rm mat}=\sqrt{2}G_F n_e\begin{pmatrix}
1+\epee & \epsilon_{e\mu}^\ast & \epsilon_{e\tau}^\ast \\
\epem & \epmm& \epsilon_{\mu\tau}^\ast \\
\epet & \epmt & \eptt \\
\end{pmatrix}~
\label{eq:ham}
\end{equation}
(up to an irrelevant constant).  Here $n_e$ is the number density of
electrons in the medium and the epsilon terms represent the contribution of the NSI.

The physical origin of the NSI terms in Eq. (\ref{eq:ham}) can be
the exchange of a new heavy vector or scalar \footnote{
Scalar
  interactions of the form $(\bar{\nu}f_R)(\bar{f}_R\nu)$ reduce to
  Eq.~(\ref{eq:lagNSI}) upon the application of the Fierz
  transformation, see e.g. S.~Bergmann, Y.~Grossman and E.~Nardi,
Phys.\ Rev.\ D {\bf 60}, 093008 (1999)
[arXiv:hep-ph/9903517].
}
particle, described by
the effective low-energy four-fermion Lagrangian
\begin{eqnarray}
L^{NSI} &=& - 2\sqrt{2}G_F (\bar{\nu}_\alpha\gamma_\rho\nu_\beta)
(\epsilon_{\alpha\beta}^{f\tilde{f} L}\bar{f}_L \gamma^\rho
\tilde{f}_L + \epsilon_{\alpha\beta}^{f\tilde{f}
R}\bar{f}_R\gamma^\rho \tilde{f}_{R})\nonumber\\
&+& h.c. \label{eq:lagNSI}
\end{eqnarray}
Here $\epsilon_{\alpha\beta}^{f\tilde{f} L}$
($\epsilon_{\alpha\beta}^{f\tilde{f} R}$) denotes the strength of
the NSI between the neutrinos $\nu$ of flavors $\alpha$ and
$\beta$ and the left-handed (right-handed) components of the
fermions $f$ and $\tilde{f}$. The epsilons in Eq.~(\ref{eq:ham})
are \emph{the sum} of the contributions from electrons
($\epsilon^{e}$), up quarks ($\epsilon^{u}$), and down quarks
($\epsilon^{d}$) in matter: $\epsilon_{\alpha\beta}\equiv
\sum_{f=u,d,e}\epsilon_{\alpha\beta}^{f}n_f/n_e$. They are thus
normalized \emph{per electron}. In turn,
$\epsilon_{\alpha\beta}^{f}\equiv\epsilon_{\alpha\beta}^{fL}+\epsilon_{\alpha\beta}^{fR}$
and
$\epsilon_{\alpha\beta}^{fP}\equiv\epsilon_{\alpha\beta}^{ffP}$.
Notice that the matter effects are sensitive only to the
interactions that preserve the flavor of the background fermion
$f$ (required by coherence \cite{Friedland:2003dv}) and,
furthermore, only to the vector part of that interaction.

Neutrino scattering tests, like those of NuTeV \cite{Zeller:2001hh} and CHARM
\cite{Vilain:1994qy}, mainly constrain the NSI couplings of the muon neutrino,
e.g., $|\epem|\lesssim 10^{-3}$, $|\epmm|\lesssim 10^{-3}-10^{-2}$. The
limits they place on $\epee$, $\epet$, and $\eptt$ are rather loose,
e. g., $|\epsilon_{\tau\tau}^{uu R}|<3$, $-0.4<\epsilon_{ee}^{uu
  R}<0.7$, $|\epsilon_{\tau e}^{uu}|<0.5$, $|\epsilon_{\tau
  e}^{dd}|<0.5 $ \cite{Davidson:2003ha}. Stronger constraints exist on
the corresponding interactions involving the charged leptons.
Those, however, are model-dependent and do not apply if the NSI
come from the underlying operators containing the Higgs fields
\cite{Berezhiani:2001rs}. Here we are interested only in direct
bounds that follow from experiments.

Given the above bounds we will set $\epem$ and $\epmm$ to zero in our
analysis. Furthermore, for simplicity we will also set $\epmt$ to
zero. The earlier analyses of the atmospheric neutrino data
\cite{Fornengo:2001pm} have indicated that this parameter is quite
constrained ($\epmt<10^{-2}-10^{-1}$).
Here, as in our previous works, we have a
three-dimensional NSI parameter space, spanned by $\epee$, $\epet$,
and $\eptt$.

\subsection{Atmospheric neutrinos and NSI}
\label{sect:atmreview}

The atmospheric neutrino flux is comprised of both neutrinos and
antineutrinos, initially produced in the electron and muon flavor
states, with the energy spectrum spanning roughly five orders of
magnitude, from $10^{-1}$ to $10^4$ GeV \cite{Engel:1999zq}.  The
data can be divided into a high energy sample, $E_\nu\gtrsim 10$
GeV, where only the muon neutrino flux is measured (through-going
and stopping muons sample), and a lower energy one, where both the
muon and electron neutrino data are available.  For the first, the
data show  disappearance of $\numu$ with large suppression
at lower energies/longer baselines and smaller suppression in the
opposite limits; the second sample is consistent with no
oscillations of $\nue$ and $\numu$ disappearance with large
amplitude.  This means that the $\numu$ oscillate into $\nutau$ at
low energy, and that the $\numu$  data are well described by
vacuum oscillations with large mixing at all energies.

One way to fit these data is to require that the NSI couplings of the
tau neutrino are negligible at all energies. For example, if
$\epet=0$, we would need $\sqrt{2}\eptt G_F n_e \lta \Delta m^2/(2
E_0)$ at $E_0 \sim 20-30$ GeV (the highest energy where an
oscillation maximum occurs for neutrinos traveling along the
diameter of the Earth), to ensure that $\numu$ oscillates into
$\nutau$ with unsuppressed amplitude in the high energy data
sample. Substituting numerical values, this translates into the
bound $|\eptt|\lesssim 0.2$ \cite{Friedland:2004ah}.

In \cite{Friedland:2004ah} we pointed out another possibility: one
of the two (non-zero) eigenvalues of $H_{mat}$ is small with
respect to the vacuum terms, so that at high energy $\numu$ can
oscillate into the corresponding eigenstate, $\nutau^\prime$,
without suppression. Specializing to the case which is
continuously connected to the standard  $\numu - \nutau$
oscillation scenario (i.e., $\nutau^\prime$ becomes $\nutau$ in
the limit of no NSI), this gives:
\begin{equation}
  \label{eq:width}
  |1+\epee+\eptt-\sqrt{(1+\epee-\eptt)^2+4 |\epet|^2}|\lesssim 0.4~.
\end{equation}
Although at high energy $\numu$'s oscillate into a mixture of
$\nutau$ and $\nue$, and not purely into $\nutau$, this is not observable,
given the absence of $\nue$ data in the high energy sample.  At lower
energy, where the kinetic terms of the Hamiltonian dominate, the
$\nue$ flux is unoscillated, as the muon neutrinos oscillate
dominantly into $\nutau$.

The condition (\ref{eq:width}) describes the allowed region of the
epsilons; it has the shape of a parabola centered on the curve
\begin{equation}
  \label{eq:parabola}
  \eptt = |\epet|^2/(1+\epee)~.
\end{equation}
It is also independent of the phase/sign of $\epet$.

The conversion of $\numu$ at high energy mimics vacuum
oscillations with an effective mass splitting, $\Delta  m^2_m$,
and an effective mixing angle, $\theta_m $, given by
\cite{Friedland:2004ah,Friedland:2005vy}:
 \beq
  &&\Delta m^2_{m} = \Delta m^2 \left[
  (c_{2\theta} (1+ c^2_\beta) - s^2_\beta)^2/4 + (s_{2\theta}
  c_{\beta})^2\right]^{1/2} ~, \nonumber \\
  &&\tan 2\theta_m = 2
  s_{2\theta} c_\beta/(c_{2\theta} (1+ c^2_\beta) - s^2_\beta )~.
 \label{effmix}
 \eeq
They depend on the NSI only through the angle $\beta$, defined as
\be
\tan 2\beta = 2 | \epet|/( 1+\epee -\eptt)~.
\label{tanb2eta}
\ee

If NSI are present but ignored in the data analysis, the highest
energy neutrino data would give $\theta_m$ and $\Delta m^2_{m}$
instead of the vacuum parameters. Those can be inferred from Eq.
(\ref{effmix}). Taking the experimentally favored value $\theta_m=
\pi/4$,  we find:
 \beq
 \Delta m^2 \simeq \Delta m^2_m(1+\cos^{-2}\beta)/2~, \hskip 0.5truecm
 \sin^2 \theta =  \frac{\Delta m^2_m}{2 \Delta m^2 }~.
 \label{cecicurve}
 \eeq
We see that the vacuum mass splitting is larger than the measured
one, $\Delta m^2_m$, and  the vacuum mixing $\theta$ is
smaller than maximal ($\theta<\pi/4$).

The vacuum parameters are constrained by data from lower energy
atmospheric neutrinos, $E \lta 1$ GeV, for which the vacuum term
of the Hamiltonian dominates. Calling $\theta_{min}$ and $\Delta
m^2_{max}$ the smallest mixing and the largest mass splitting
allowed by those, we have two constraints on $\beta$
\cite{Friedland:2005vy}:
 \beq
  \cos^2 \beta \gta \tan^2 \theta_{min}~, \hskip 0.5truecm
  \cos^2 \beta \geq \left[\frac{2 \Delta m^2_{max}}{\Delta m^2_m}-1 \right]^{-1}~,
 \label{ceci&alexcond}
  \eeq
that follow from Eq. (\ref{cecicurve}). These conditions truncate
the  allowed region (\ref{eq:width}), as shown in
\cite{Friedland:2004ah,Friedland:2005vy}.

Let us emphasize that the analytics in Eqs.
(\ref{effmix})-(\ref{ceci&alexcond}) is for guidance only, since
it relies on the over-restrictive condition (\ref{eq:parabola})
and on the approximation of negligible vacuum terms at $E \gta 10$
GeV. The corrections due to the vacuum terms to the conditions
(\ref{eq:width}) and (\ref{ceci&alexcond}) introduce a dependence
on the mass hierarchy, i.e. on the sign of $\dm$, with the
inverted hierarchy ($\dm<0$) giving a larger allowed region of the
NSI.  We refer to ref. \cite{Friedland:2005vy} for details on this
and on the corrections due to the mixing angle $\theta_{13}$,
which break the symmetry in the phase of $\epet$.


\section{$\numu -\nutau$ oscillations at MINOS}
\label{oscillations}

\subsection{General considerations for K2K and MINOS}

For the setups of K2K and MINOS the neutrino oscillations in the
$\numu -\nutau$ channel are dominated by the vacuum contribution, with
the matter effects playing only a subdominant role.  This is proven by
the fact that both beams have baselines much shorter than the
characteristic length (or equivalently, the width, see
\cite{Lunardini:2000sw}) which describes refraction effects, $l_{ref}
= (\sqrt{2} G_F n_e)^{-1} \simeq 1.9 \cdot 10^{3} $ Km for the average
density of the continental crust, $\rho = 2.7 ~{\rm g \cdot cm^{-3}}$.
Specifically, the K2K baseline is 250 Km long, only about a tenth of
the refraction length, and this ensures that matter effects at K2K are
negligible.  MINOS, with its 735 Km long baseline, covers about 40\%
of $l_{ref}$, and therefore is expected to see matter effects at a
subdominant, while not entirely negligible, level.  For simplicity
here we consider the zeroth order description of both K2K and MINOS in
terms of vacuum oscillations only (and comment on the small corrections
due to refraction effects at MINOS later, see sec. \ref{results}).




In this approximation MINOS and K2K measure the oscillation parameters
in vacuum, $\theta$ and $\Delta m^2$. The difference between
$\theta, \Delta m^2$ and the corresponding parameters inferred
from the atmospheric neutrino data tells us how much room there is
for NSI, through Eqs. (\ref{cecicurve})-(\ref{ceci&alexcond}).

The present MINOS data (a similar description holds for K2K) can
be fit by a range of $\Delta m^2$ and $\theta$, such that larger
values of $\Delta m^2$ come with smaller values of $\theta$.
Compared to the corresponding region from the SuperKamiokande
atmospheric analysis, the MINOS region is relatively narrow in
$\Delta m^2$, $\Delta m^2 = (2.5 - 3.65) \unitdm $ (at 68\%
confidence level, C.L.) \cite{talkpetyt}, but extended in
$\theta$, with $\sin^2 2\theta \simeq 0.73 - 1 $ (68\% C.L.).
The
MINOS constraint on $\Delta m^2$ also improves slightly on the
analogous restriction given by K2K.

We notice that the  region currently allowed by MINOS is
extended in the direction (smaller $\theta$ comes with larger
$\Delta m^2$)
obtained from the NSI analysis of the atmospheric data.
Due to this accidental degeneracy,
we expect that the present-day MINOS results leave substantial
room for the NSI, with only a loose upper bound on $\beta$
according to Eq. (\ref{ceci&alexcond}).


\subsection{The data analysis: MINOS, SuperKamiokande, K2K}

We have carried out a combined analysis of the atmospheric data
from SuperKamiokande and $\nu_\mu\rightarrow\nu_\tau$ oscillation
data from K2K and MINOS. Our study has two parts. First, we have
investigated the impact of the recently released MINOS data on the
existing NSI constraints. Second, to illustrate the expected
future sensitivity of MINOS, we have simulated a future MINOS signal
with projected higher statistics. For this simulation, we
considered two scenarios: one which would give a signal of NSI, and
another that would instead give  a constraint on them.

The starting point in both cases are the results from our NSI
analysis of the atmospheric neutrino data. This analysis was done
in \cite{Friedland:2004ah} and \cite{Friedland:2005vy} and
involved a fit to the complete 1489-day charged current
Super-Kamiokande phase I data set
\cite{Hayato:2004aw,Ashie:2005ik}, including the $e$-like and
$\mu$-like data samples of sub- and multi-GeV contained events as
well as the stopping and through-going upgoing muon data events.
The details of the code,  provided by Michele Maltoni, are given
in \cite{Friedland:2004ah,Friedland:2005vy}. In
\cite{Friedland:2005vy} we have scanned a five-dimensional
parameter space $(\epee,\epet,\eptt,\Delta
m_{23}^2,\sin^2\theta_{23})$ and obtained an allowed region in
this space. The output of that scan -- a set of $\chi^2$ values on
a five-dimensional grid -- was used here.

The output was combined with a $\chi^2$ array from the K2K vacuum
oscillation analysis. The latter was provided by the K2K
collaboration in tabular form, and refers to the published
analysis of Ref. \cite{Aliu:2004sq} (Fig. 4 there). Since that
analysis is practically independent of the matter interactions,
the same $\chi^2$ array was added to each $(\Delta
m_{23}^2,\sin^2\theta_{23})$ plane. For details, see again
\cite{Friedland:2005vy}.

Finally, we have performed a similar fit to the released MINOS
data. The data used were the number of observed events in each
MINOS energy bin versus the corresponding predictions without
oscillations as given in \cite{talkpetyt}. Our analysis
follows the one in \cite{talkpetyt} as closely as possible.
The resulting allowed region and the best-fit point agree
reasonably well with those in the MINOS analysis
(\cite{talkpetyt}, slide 60). The output of this fit was then
combined with those of the  SuperKamiokande and K2K
analyses, giving a five-dimensional allowed region in the space
$(\epee,\epet,\eptt,\Delta m_{23}^2,\sin^2\theta_{23})$. The
combined results were then marginalized, either over the NSI or
over the oscillation parameters.

For the second part, we have simulated the data with increased
statistics that might be expected in the future at MINOS. We have
used statistics corresponding to $25\times 10^{20}$
protons-on-target (p.o.t.), as quoted by the MINOS collaboration
(see, {\it e.g.}, \cite{talkpetyt}), henceforth referred to as
``MINOS-high'' sample. We have also checked how our results change
with changing the statistics to $16\times 10^{20}$ p.o.t.

We have simulated future data by taking the histogram of the expected
number of events in each energy bin  published  by the MINOS
collaboration, scaling it up to the appropriate statistics,
applying oscillations to it according to the set of parameters
chosen and finally simulating statistical fluctuations in the data
using Poisson statistics. It is hoped that this simplified
procedure  still captures the relevant effects. The most accurate
way would of course be to perform a full Monte-Carlo simulation of
the detector, which could be done by the MINOS collaboration.

The simulated data were then used as an input to the combined
SuperKamiokande-K2K-MINOS analysis described before.


\subsection{Results with present MINOS data}
\label{results}

\begin{figure}[htbp]
  \centering  \includegraphics[width=0.35\textwidth]{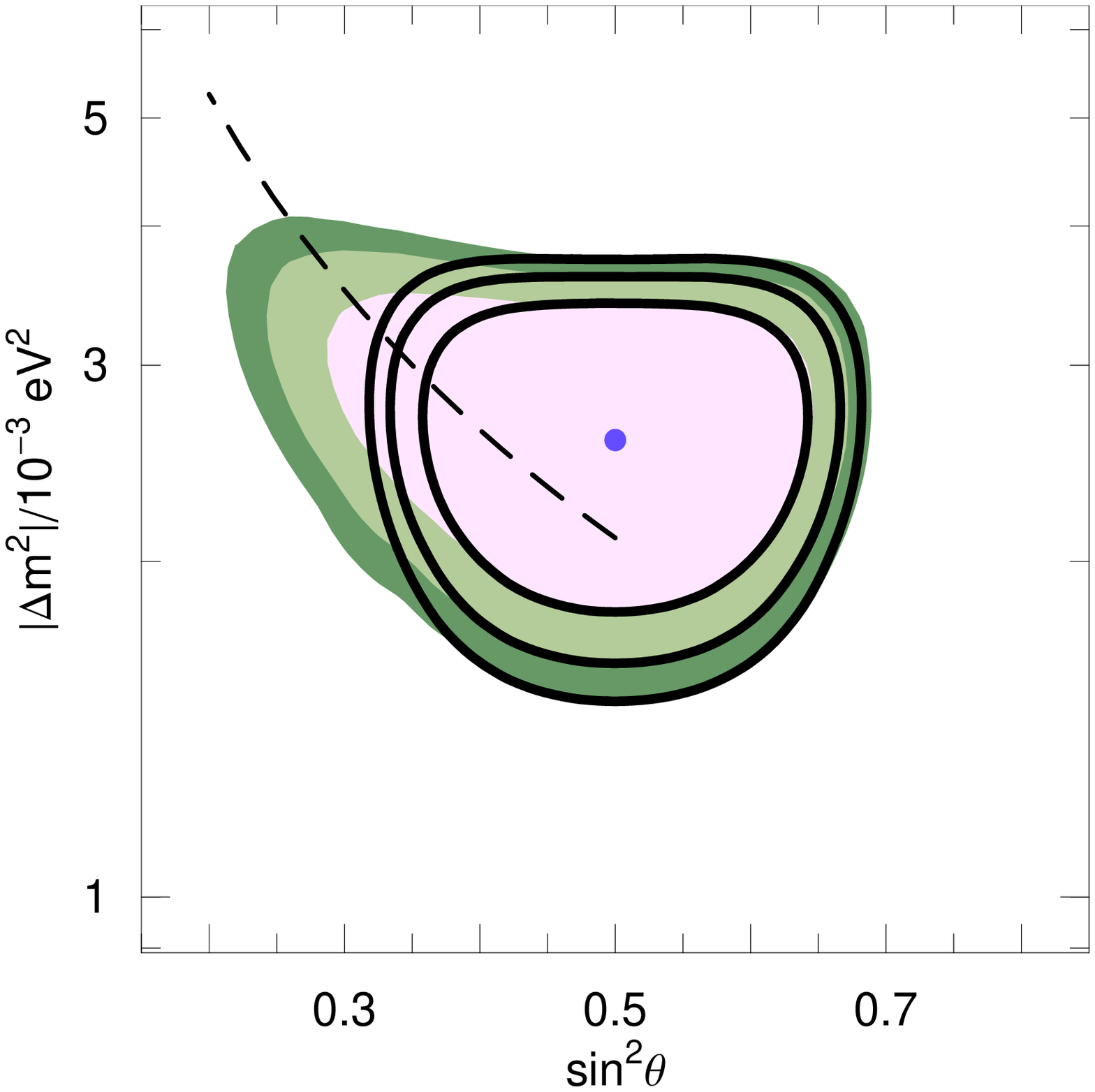}
  \includegraphics[width=0.35\textwidth]{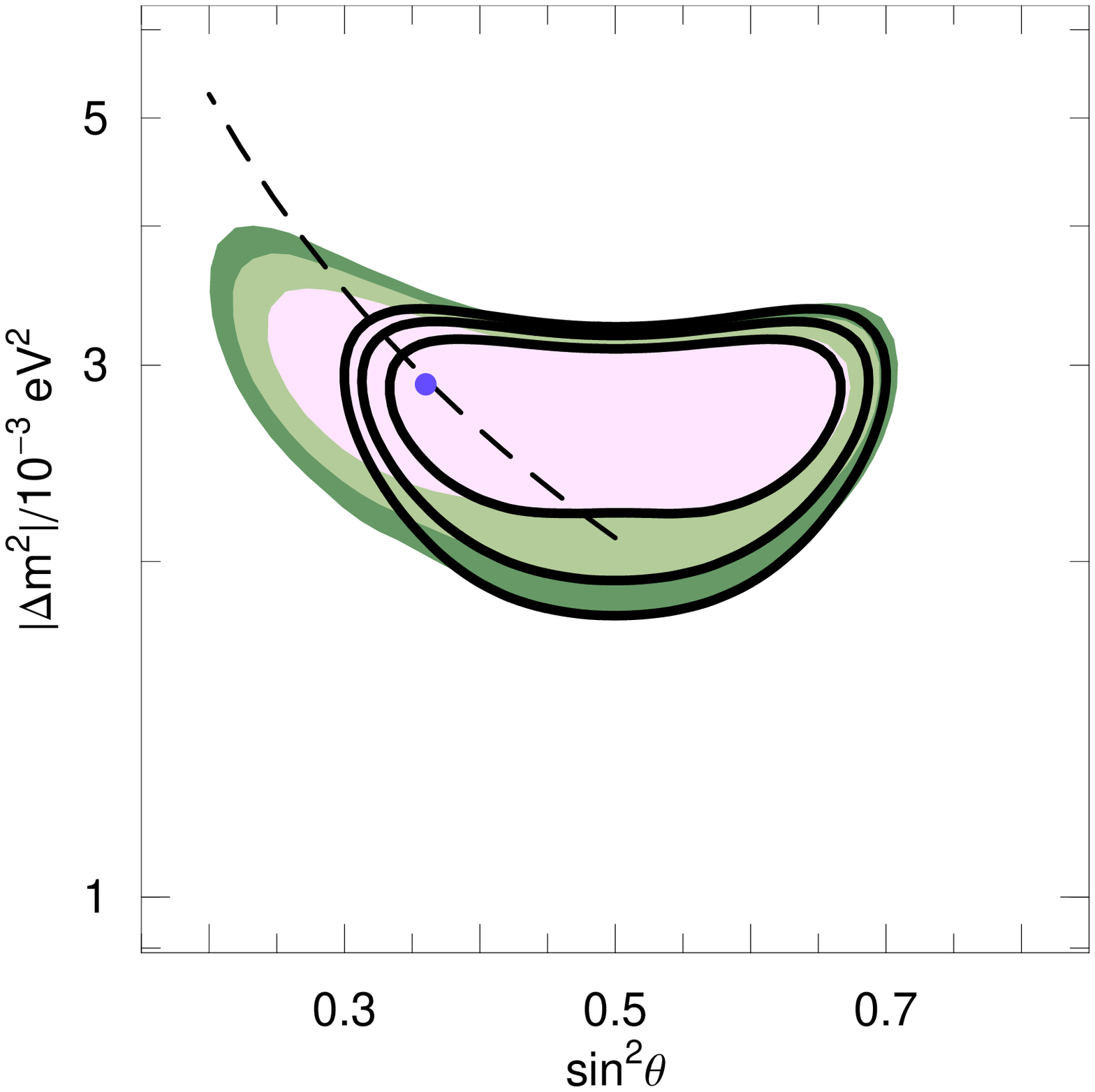}
  \caption{Regions in the space $\dm - \sin^2 \theta$ allowed by the
  global fit before (upper panel) and after (lower panel) the MINOS
  results, with purely standard interactions (contours) and with NSI
  (filled areas).  For both cases we plot the regions allowed at 95\%,
  99\% and $3\sigma$ confidence levels for 2 degrees of freedom.  We
  have marginalized also over the sign of $\Delta m^2$ and took
  $-1\leq \epee \leq 1.6$, motivated by one of the accelerator
  bounds (see \cite{Friedland:2004ah}). The dashed line represents the possible
  positions of the best fit point of the atmospheric data only in the
  space of the vacuum parameters for fixed $\theta_m$ and $\Delta
  m^2_m$, and varying NSI along the parabola (\ref{eq:parabola}).  We
  used $\Delta m^2_m = 2.1 \cdot 10^{-3}~{\rm eV^2}$ and
  $\theta_m=\pi/4$, motivated by the best fit of atmospheric neutrinos
  alone \cite{Ashie:2005ik}.
  }
  \label{greenfigure}
\end{figure}

\begin{figure}[htbp]
 \centering
  \includegraphics[width=0.35\textwidth]{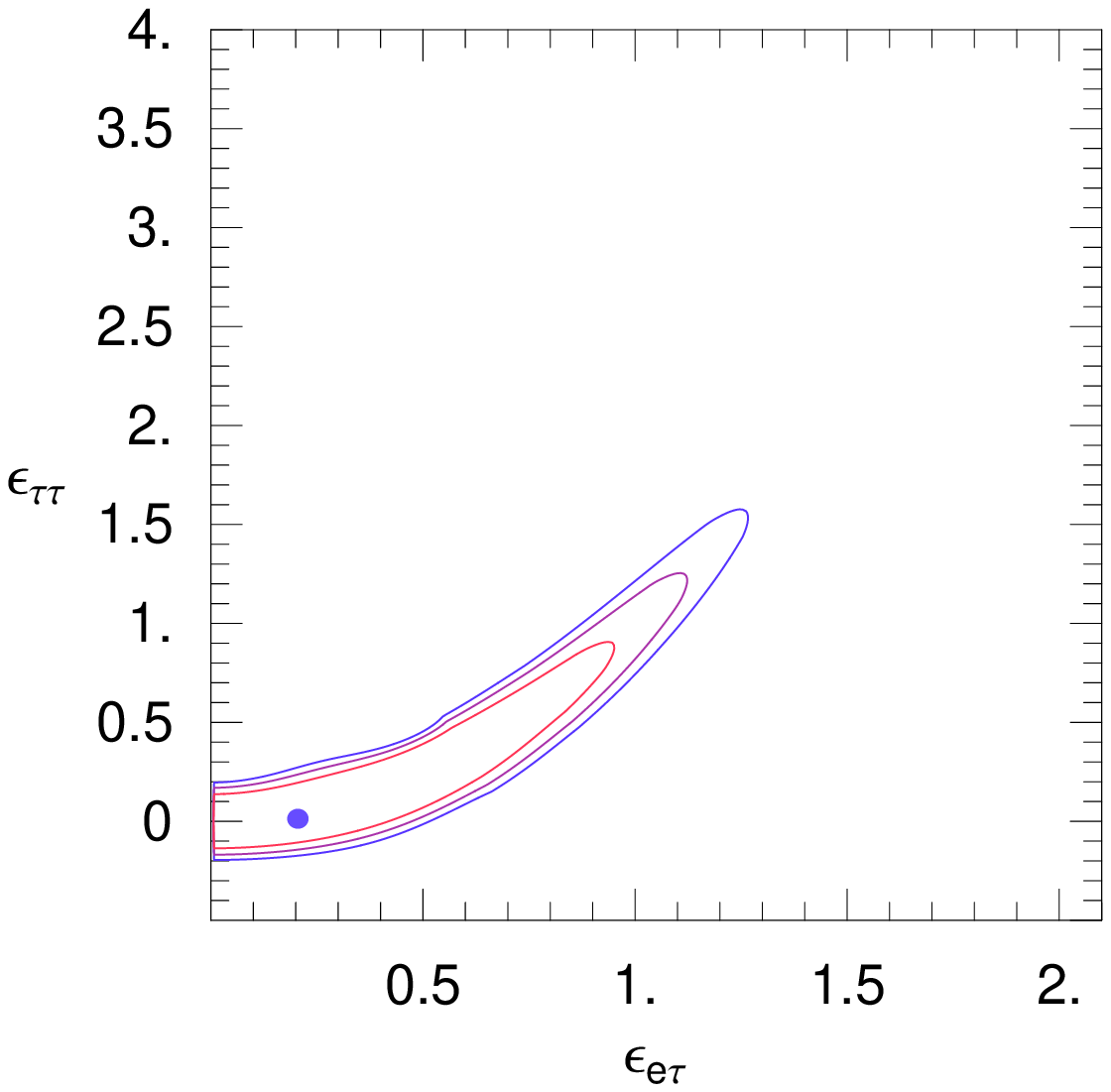}
 \includegraphics[width=0.35\textwidth]{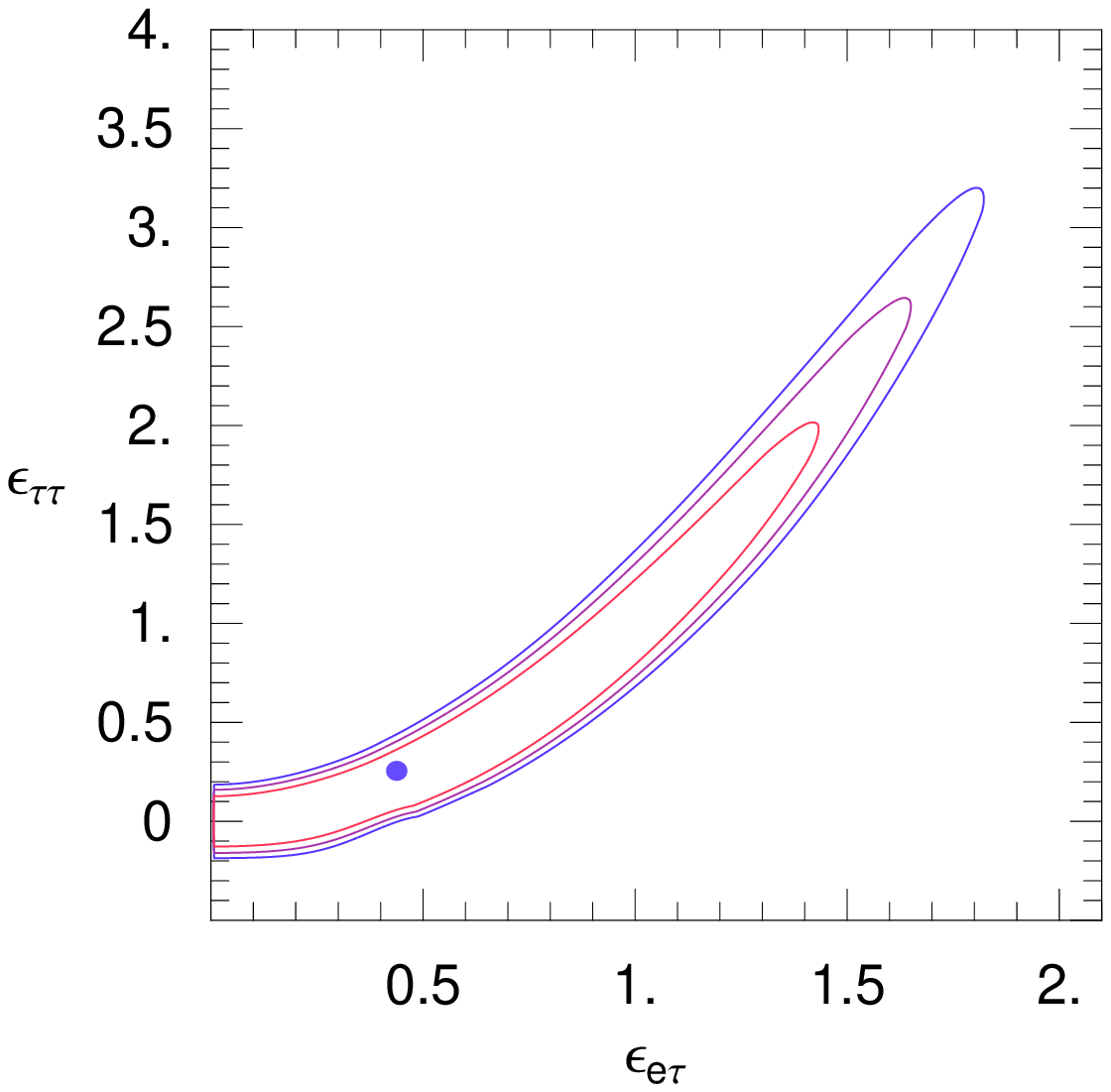}
  \caption{Two sections of the allowed region in the three dimensional
  space of the NSI couplings along the plane $\epee= 0$. The contours
  refer to 95\%, 99\% and $3\sigma$ confidence levels.  The points of
  minimum of the $\chi^2$ in this plane are also shown. The upper
  (lower) panel refers to normal (inverted) mass hierarchy. The
  $\epee<0$ regions are symmetric with respect to those shown. }
  \label{smiles}
\end{figure}

The results of our analysis of the existing data are presented in
Figs. \ref{greenfigure}-\ref{smiles} and Table \ref{summarytable}.
%

Fig. \ref{greenfigure} shows the effect of adding the MINOS
results on the allowed region in the space of $ \sin^2 \theta -
\Delta m^2 $. The contours show the case of the standard
interactions, while the filled regions were found after marginalizing
over the NSI (see the caption for details). The Figure also shows
the direction in Eq. (\ref{cecicurve}), along which $ \sin^2
\theta$ and $ \Delta m^2 $ would be found for fixed $ \sin^2
\theta_m$ and $ \Delta m^2_m $ ($\Delta m^2_m = 2.1 \cdot
10^{-3}~{\rm eV^2}$ and $\theta_m=\pi/4$, motivated by the best
fit to the atmospheric data only \cite{Ashie:2005ik}) and NSI
varying along the parabola (\ref{eq:parabola}).

For the case of no NSI we see that both with and without MINOS the
best fit point lies at maximal mixing, with the interval $\sin^2
\theta \simeq 0.3 - 0.7$ allowed at $3 \sigma$ confidence level. These
bounds on the mixing are due to the atmospheric data; the main
effect of adding the MINOS data is to further restrict $\Delta
m^2$: $\Delta m^2= (1.8 - 3.4)\unitdm $ at $3 \sigma$ C.L.

Both with and without MINOS, in the case with NSI the allowed
region extends along the direction in Eq. (\ref{cecicurve}),
allowing smaller mixing and larger $\Delta m^2$ with respect to
the standard case. Besides the restriction on $\Delta m^2$ already
observed, adding the MINOS data has the effect to move the best
fit point away from maximal mixing. The new best fit point has
$\sin^2 \theta=0.36$ and $\dm = 2.88 \unitdm$, and it lies along
the curve (\ref{cecicurve}) as it is plotted in the figure.  This
happens because the MINOS-only best fit point -- $\sin^2
\theta=0.33$ and $\dm = 3.05 \unitdm$ \cite{talkpetyt} -- lies
near this curve and the atmospheric $\chi^2$ is very flat along
this direction, once the NSI are included.  It has to be stressed,
however, that there is only an insignificant difference between
the best fit point and, say, the minimum of $\chi^2$ along the
direction $\theta=\pi/4$.  The latter point is within the $68\%$
C.L. contour.

The shift of the best fit point in the space of oscillation
parameters indicates that with MINOS non-zero NSI are favored,
even though in a statistically insignificant way.  To check this,
we have marginalized the likelihood over the oscillation
parameters (with fixed mass hierarchy) to obtain the $\chi^2$ in
the three dimensional space of the epsilons,
$\chi^2(\epee,\eptt,\epet)$.  We find that for the inverted mass
hierarchy the minimum of this marginalized $\chi^2$ is at the
point $(\epee,\eptt,\epet)=(0.6,0.933,0.615)$. This point is
practically degenerate with others that satisfy the condition
(\ref{eq:parabola}) and correspond to vacuum oscillation
parameters (Eq. (\ref{cecicurve})) close to the MINOS-only best
fit point. Again, we find that the difference in $\chi^2$ between
the best fit point and the point with zero NSI is minimal and has
no statistical significance.

In Fig. \ref{greenfigure} the effects of matter on the MINOS
signal were not taken into account for simplicity.  We have
checked that their inclusion leaves the results unchanged
qualitatively, with only a change of the order of 10\% in the
extent of the upper left part of the allowed region (for the NSI
case, shaded areas).

Fig. \ref{smiles} shows sections of the 3D allowed region along the
plane $\epee=0$  for normal and inverted mass hierarchy.  The
contours correspond to intervals of $\chi^2$ with respect to the
absolute minimum given above and represent the regions allowed at
95\%, 99\% and $3\sigma$ confidence levels.  They show only minor
changes with respect to the same contours without MINOS \cite{Friedland:2005vy}.  For
illustration in Fig. \ref{smiles} we show the points of minimum of
$\chi^2$ along the plane $\epee=0$.  Those appear to be away from the
origin, confirming what has been discussed about the shift of the best
fit to non-zero NSI.



\begin{table}
\begin{tabular}{| l | l | l |l |}
\hline
\hline
$\epee $ & $|\epet|$ (99\% C.L.) & $\eptt$ (99\% C.L.)& $\epet,\eptt$ (best) \\
\hline
\hline
-1  & $<$0.3 & $\left[ -0.4 ,  0.6\right]$  & 0, 0\\
\hline
-0.5  & $<$0.75 & $\left[-0.15 , 1.2 \right]$ & 0.15, 0.1\\
\hline
0.  & $<$1.65 & $\left[-0.15 , 2.6 \right]$ & 0.438,  0.255\\
\hline
0.3  & $<$2.1 & $\left[-0.15 , 3.4 \right]$ & 0.67,  0.413\\
\hline
0.6  & $<$2.5 & $\left[-0.15 , 3.9 \right]$ & 0.933,  0.615 \\
\hline
0.9  & $<$2.9  & $\left[-0.15 , 4.5 \right]$ & 1.019,  0.615 \\
\hline
\hline
\end{tabular}
\caption{Some relevant numbers describing the 3D allowed region of
the NSI couplings for inverted mass hierarchy.  We give the
intervals of $\epet$ and of $\eptt$ enclosed by the 99\% C.L.
region in planes of fixed $\epee$, and the values of $\epet,\eptt$
that minimize the $\chi^2$ in the same planes.  }
\label{summarytable}
\end{table}

Table \ref{summarytable} gives further details on our results.  It
presents information on sections of the 3D allowed region along
several planes of fixed $\epee$, thus generalizing what is shown in
Fig. \ref{smiles}.

\subsection{Expected future sensitivity of MINOS}

\begin{figure}[htbp]
  \centering
  \includegraphics[width=0.35\textwidth]{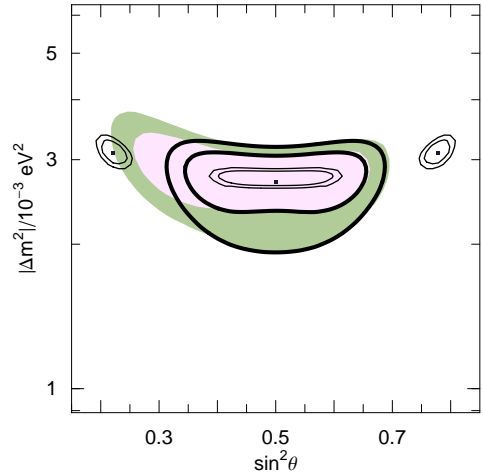}
  \caption{Results of fits to simulated MINOS data with statistics
increased from the current $0.93 \cdot 10^{20}$ to $25 \cdot
10^{20}$ protons on target (thin contours). 90\% and 99\% C.L.
regions are shown. The ``data" were simulate for two sets of NSI
and ``true'' oscillation parameters: (i) no NSI, $\sin^2
\theta=0.5$ and $ \Delta m^2=2.7 \unitdm$, (ii) $\epee=0$,
$\eptt=0.81$, $\epet=0.9$, $\sin2 \theta=0.27$ and $ \Delta
m^2=3.1 \unitdm$. The fits were done in both cases in the
assumption of no NSI. For reference, we also show the regions
allowed currently by all the data combined, at 90\% and 99\% C.L.
with (filled area) and without NSI (thick contours), as in Fig.
\ref{greenfigure}. See text for details.
}
  \label{fakefits}
\end{figure}

How will the constraints on NSI improve with the increase of the
statistics at MINOS?  We consider the MINOS-high setup for two
examples of NSI and ``true'' oscillation parameters: (i) no NSI,
$ \sin^2 \theta=0.5$ and $ \Delta m^2=2.7 \cdot 10^{-3}{\rm eV^2} $,
and (ii) $\epee=0$, $\eptt=0.81$, $\epet=0.9$,  $ \sin^2 \theta=0.27$
and $ \Delta m^2=3.1 \cdot 10^{-3}{\rm eV^2} $.
Fig. \ref{fakefits} gives the results of our fit to simulated
MINOS data.
In both cases (i)  and (ii), the fits were done \emph{under the
assumption of no NSI} \footnote{Thus, in presence of NSI the
parameters measured in this way are the mass splitting and mixing
in matter, $\theta_m$ and $\Delta m^2_m$, which for the MINOS
setup differ from $\Delta m^2_m \simeq \Delta m^2$ only in small
corrections}. The aim was to see if a large NSI scenario would
result in a strong tension between the MINOS data and the
atmospheric neutrino fits, if both were analyzed assuming standard
matter interactions. Indeed, this happens for scenario (ii): the
high statistics MINOS allowed region has no overlap with the
current region from MINOS+atmospheric+K2K. This way, the MINOS
measurement would strongly favor the presence of NSI versus the
pure standard interaction. 
Once the NSI are included in the fit, the tension disappears.
We find that the allowed region in the
space of the epsilons (marginalized over the vacuum parameters)
includes the case of standard interactions, $\epee=\epet=\eptt=0$,
only with confidence level higher than 99\%.

Note that in both cases shown in  Fig. \ref{fakefits}, $\Delta
m^2$ would be measured with approximately 6\% precision, and
$\sin^2 \theta$ with 15\% precision. Notice also that, for case
(ii), the MINOS-high allowed region does not include the input
values used for the vacuum parameters. This indicates that the
small corrections due to matter effects on the MINOS beam are
sufficiently important for the precision of MINOS-high.


We obtain very similar results with a more modest increase of the
MINOS statistics, to $16 \cdot 10^{20}$ instead of $25 \cdot
10^{20}$ protons on target. With this intermediate increase, the
point $\epee=\eptt=\epet=0$ is inside the 99\% C.L. region, but
outside the 95\% C.L. contour.

\section{NSI-driven $\nu_\mu\rightarrow\nu_e$ conversion at MINOS}
\label{numunue}

So far, we have discussed the dominant
$\nu_\mu\rightarrow\nu_\tau$ oscillation mode and argued that the
effects of intervening matter on this mode are subdominant. We next show that for the
$\nu_\mu\rightarrow\nu_e$ mode matter effects due to NSI can be
instead very important. This opens up another, complementary mode for
searching for NSI at MINOS.

The $\nu_\mu\rightarrow\nu_e$ mode is being studied by MINOS as a
way to measure/constrain the  $\theta_{13}$ mixing angle.
According to the analysis by the collaboration, with the planned higher
statistics MINOS will have an impressive sensitivity to
$\theta_{13}$ in this mode, providing a significant improvement
over the current CHOOZ bound. For instance, with $25\times10^{20}$
protons on target and assuming $\Delta m_{23}=2.5\times10^{-3}$
eV$^2$ the MINOS bound is projected to be $\sin^2
2\theta_{13}\lesssim0.07$, compared to $\sin^2
2\theta_{13}\lesssim0.14$ as currently given by CHOOZ.

If the non-standard flavor changing interaction $\epsilon_{e\tau}$
is present, it will also drive $\nu_\mu\rightarrow\nu_e$
conversion. Schematically, this conversion can be viewed in two
steps:
\begin{equation}\label{eq:twostep}
    \nu_\mu\stackrel{\Delta_{23},\theta_{23}}{\longrightarrow}\nu_\tau
    \stackrel{\epsilon_{e\tau}}{\longrightarrow}\nu_e.
\end{equation}
The first step has already been observed by MINOS, with the largest
conversion happening in the lower energy part of its spectrum ($1.5-2$
GeV). Correspondingly, the $\nu_e$ production according to
Eq.~(\ref{eq:twostep}) is also expected to peak at low energy.  Below,
we quantify this by giving an approximate analytical expression for
the probability $P(\nu_\mu\rightarrow\nu_e)$ and presenting a sample
numerical calculation.

\subsection{Analytical treatment of $\nu_\mu\rightarrow\nu_e$ conversion}

We begin by noting that the probability
$P(\nu_\mu\rightarrow\nu_e)$ will be small $(\ll1)$. This will
allow us to treat the process perturbatively. To introduce the
idea, let us first consider a toy example. Suppose we have a
two-level problem with the Hamiltonian
$$H_{2\nu}=\begin{pmatrix}
 \delta_1 & \delta_{12} \\
 \delta_{12} & \delta_2
 \end{pmatrix}$$
and the initial state is $(0,1)^T$. Assume further that
$|\delta_{12}|\ll|\delta_2-\delta_1|$. Then the probability
$P(2\rightarrow1)$ can be computed in perturbation theory as
\begin{eqnarray}\label{P2nuapprox}
    P(2\rightarrow1)&\simeq& \left|\int_0^t dt' (-i)
\delta_{12}e^{-i(\delta_2-\delta_1)t'}\right|^2\nonumber\\
&=&\left|
\delta_{12}\frac{1-e^{-i(\delta_2-\delta_1)t}}{(\delta_2-\delta_1)}\right|^2.
\end{eqnarray}
To arrive at this one can (i) subtract $\delta_1$ on the diagonal
(overall constants on the diagonal do not change oscillation
probabilities), (ii) solve the evolution equation for $\nu_2$
neglecting the effects of $\nu_1$, $i\partial_t \psi_2\simeq
(\delta_2-\delta_1)\psi_2$, and finally (iii) solve the equation
for $\nu_1$ treating $\nu_2$ obtained before as a source,
$i\partial_t \psi_1= \delta_{12}\psi_2$. The second step is the
one containing an approximation, it assumes that
$|(\delta_2-\delta_1)\psi_2|\gg|\delta_{12}\psi_1|$.

It is useful to compare Eq.~(\ref{P2nuapprox}) with the exact
solution,
\begin{equation}
    P(2\rightarrow1) =\left|\delta_{12}
    \frac{1-e^{-i\sqrt{(\delta_2-\delta_1)^2+4\delta_{12}^2}t}}
    {\sqrt{(\delta_2-\delta_1)^2+4\delta_{12}^2}} \right|^2.
\end{equation}
This explicitly shows that (\ref{P2nuapprox}) is a good
approximation so long as $(\delta_2-\delta_1)^2\gg4\delta_{12}^2$.

Next, consider a three-level problem with
$$H_{3\nu}=\begin{pmatrix}
 0 & \delta_{12} & \delta_{13}\\
 \delta_{12} & \Delta_2 & 0 \\
 \delta_{13} & 0        & \Delta_1
 \end{pmatrix}$$
Notice that the 23 block has a diagonal form. Suppose the initial
state is $\nu_{23}\equiv (0,a,b)^T$ and we again want the probability of
finding the particle in the first state. If the off-diagonal
entries in this Hamiltonian are smaller than the diagonal ones, we
can generalize the previous method to give
\begin{eqnarray}\label{P3nuapprox}
    P(\nu_{23} \rightarrow1)&\simeq& \left|(-i)\int_0^t dt'
[a\delta_{12}e^{-i \Delta_2t'}+b\delta_{13}e^{-i
\Delta_1t'}]\right|^2\nonumber\\
  &=&\left|a\delta_{12}\frac{1-e^{-i\Delta_2t}}{\Delta_2}
 +b\delta_{13}\frac{1-e^{-i\Delta_1t}}{\Delta_1}\right|^2.
\end{eqnarray}

The problem we are interested in, that of $
\nu_\mu\rightarrow\nu_\tau\rightarrow\nu_e$ conversion, can be
reduced to the last case. For that, we need to diagonalize the
$\nu_\mu-\nu_\tau$ block of the Hamiltonian. This is done by
rotating this block by an angle $\theta_{23}'$ given by
\begin{eqnarray}
 \label{theta23prime}
 \tan2\theta_{23}' &=& \frac{\Delta\sin2\theta_{23}}{\Delta\cos2\theta_{23}
 +\sqrt{2}G_F N_e\epsilon_{\tau\tau}/2}.
\end{eqnarray}
Notice that this angle is modified from its standard value
$\theta_{23}$ by the presence of the NSI matter term.

The two relevant mass splittings are
\begin{widetext}
\begin{eqnarray}
 \label{Delta12}
 \Delta_{1,2} \simeq \frac{1}{2}\left( \sqrt{2} G_F N_e
 \epsilon_{\tau\tau}\pm
 \sqrt{(\sqrt{2} G_F N_e \epsilon_{\tau\tau})^2+
 4(\Delta^2+\sqrt{2} G_F N_e \epsilon_{\tau\tau} \Delta\cos2\theta_{23})}\right)+
 \Delta - \sqrt{2} G_F N_e (1+\epsilon_{ee}),
\end{eqnarray}
\end{widetext}
where $\Delta_1$ comes with the ``+" sign. Normal hierarchy is
assumed for definiteness.
The small solar splitting is neglected.

The state $\nu_\mu$ is a linear combination of the new basis
states; correspondingly, the initial wavefunctions are given by
$a=\cos\theta_{23}'$ and $b=-\sin\theta_{23}'$ in the rotated
basis. We also need to rotate the NSI $\epsilon_{e\tau}$ term and
the vacuum $\theta_{12}$ and $\theta_{13}$ terms. It is important
to stress that the $\epsilon_{e\tau}$ term is rotated from the
flavor basis (by the angle $\theta_{23}'$), while the other two
are rotated from the standard (no NSI) basis that diagonalizes the
23 block (by the angle $\theta_{23}'-\theta_{23}$).

Combining these ingredients, we finally obtain
\begin{widetext}
\begin{eqnarray}
 \label{Pmuegeneralanswer0}
  P(\nu_\mu\rightarrow\nu_e) &\simeq& \left|
  G_1
  \sin\theta_{23}'\frac{\exp(i\Delta_1 L)-1}{\Delta_1}
   -G_2
   \cos\theta_{23}'\frac{\exp(i\Delta_2 L)-1}{\Delta_2}
  \right|^2,
\end{eqnarray}
where
\begin{eqnarray}
 \label{PmuegeneralanswerG1}
  G_1 &=&
  \sqrt{2} G_F N_e \epsilon_{e\tau}\cos\theta_{23}'
  + \Delta\sin2\theta_{13}\cos(\theta_{23}'-\theta_{23})
  + \Delta_\odot\sin2\theta_{12}\sin(\theta_{23}'-\theta_{23}),\\
   \label{PmuegeneralanswerG2}
  G_2 &=&   \sqrt{2} G_F N_e \epsilon_{e\tau}\sin\theta_{23}'
   + \Delta\sin2\theta_{13}\sin(\theta_{23}'-\theta_{23})
  - \Delta_\odot\sin2\theta_{12}\cos(\theta_{23}'-\theta_{23}).
\end{eqnarray}
\end{widetext}

\subsection{Sample numerical calculation of $\nu_\mu\rightarrow\nu_e$ conversion}

\begin{figure}[htbp]
  \centering
  \includegraphics[width=0.47\textwidth]{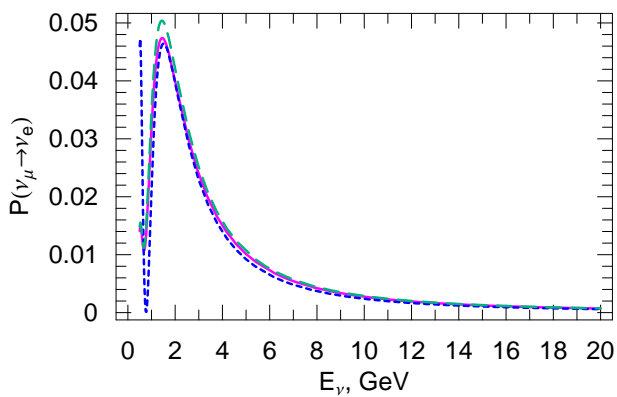}
  \caption{Conversion probability $P(\nu_\mu\rightarrow\nu_e)$ as a function of energy for
  (i) $\sin^2 2\theta_{13}=0.07$, $\Delta m_{23}=2.5\times10^{-3}$ eV$^2$,
  $\sin^2\theta_{23}=1/2$ and standard neutrino interactions (short-dashed curve), vs.
  (ii) $\sin^2 2\theta_{13}=0$, $\Delta m_{23}=2.9\times10^{-3}$ eV$^2$,
  $\sin^2\theta_{23}=0.36$ and $\epsilon_{ee}=0$, $\epsilon_{e\mu}=0.9$,
  $\epsilon_{\tau\tau}=0.81$ (solid curve). Both curves were computed numerically.
  For comparison, the predictions of the analytical result,
  Eqs.~(\ref{Pmuegeneralanswer0},\ref{PmuegeneralanswerG1},\ref{PmuegeneralanswerG2}), for the case
(ii) are also shown (long-dashed curve).}
  \label{fig:Pnumu_nue}
\end{figure}

We are now ready to explore the sensitivity of the
$\nu_\mu\rightarrow\nu_e$ conversion mode to NSI quantitatively.
As already mentioned, the MINOS experiment expects to extend the
lower bound on $\theta_{13}$ down to $\sin^22\theta<0.07$. We plot
the corresponding probability $P(\nu_\mu\rightarrow\nu_e)$ for this
case in Fig.~\ref{fig:Pnumu_nue} (short-dashed curve). We also
plot the case with $\theta_{13}=0$ and values of NSI designed to
imitate the same effect: $\epsilon_{ee}=0$, $\epsilon_{e\mu}=0.9$,
$\epsilon_{\tau\tau}=0.81$ (solid curve). Notice that the epsilons
were chosen to lie in the presently allowed region, by satisfying
Eq.~(\ref{eq:parabola}). The details of the parameters are given
in the Figure caption.

The Figure shows that the two cases, while quite different for
$E_\nu<1$ GeV, give indistinguishable signals for $E_\nu>1$ GeV,
the relevant energy range for MINOS. This fact, already described in \cite{Huber:2001de} in the context of neutrino factories, means that the 3$\sigma$
bound expected for $\theta_{13}$ could be translated into a
corresponding bound on the NSI.

Our analytical results agree very well with
those of the exact numerical calculations. This can be
seen in Fig.~\ref{fig:Pnumu_nue}, where the long-dashed curve
shows $P(\nu_\mu\rightarrow\nu_e)$ computed according to
Eqs.~(\ref{Pmuegeneralanswer0},\ref{PmuegeneralanswerG1},\ref{PmuegeneralanswerG2})
for the same choice of parameters as the solid curve discussed
above. The agreement between the two is more than adequate.

\subsection{Discussion: degeneracy of $\theta_{13}$ and NSI effects; implications for bounds}

So far we have shown that the NSI are constrained under the
assumption $\theta_{13}=0$.
The situation is more complicated if the effects of NSI and
$\theta_{13}$ are comparable. It is easy to see from
Eqs.~(\ref{Pmuegeneralanswer0},\ref{PmuegeneralanswerG1},\ref{PmuegeneralanswerG2})
that the two can interfere. The interference can be either
constructive or destructive, depending on the \emph{relative
phases} of $\epsilon_{e\tau}$ and $e^{i\delta}\sin 2\theta$. Thus,
a non-observation of the $\nu_\mu\rightarrow\nu_e$ conversion
mode at MINOS strictly speaking would not give a simple bound on
$\theta_{13}$ or $\epsilon_{e\tau}$, but would define an allowed
region in the $e^{i\delta}\sin 2\theta_{13}$ -- $\epsilon_{e\tau}$
parameter space. The degeneracy would have to broken by some other
means.

\subsection{Note on MINOS baseline advantage over K2K}

Finally, we note that MINOS, with its three times longer baseline,
possesses an important advantage in sensitivity to the NSI
relative to K2K. Indeed, consider
Eqs.~(\ref{Pmuegeneralanswer0},\ref{PmuegeneralanswerG1},\ref{PmuegeneralanswerG2})
and suppose that both the baseline $L$ and the neutrino energy $E_\nu$
($\Delta \propto 1/E_\nu$) are varied in such a way that for
shorter baseline the oscillation phase in the
$\nu_\nu\rightarrow\nu_\tau$ channel is preserved. It is then
clearly seen that the matter term, $\propto \sqrt{2} G_F N_e
\epsilon_{e\tau}$, becomes relatively less important than the
vacuum oscillation term driven by $\theta_{13}$,
$\propto\Delta\sin 2\theta_{13}$. Thus, the K2K experiment is
less sensitive to the NSI than MINOS. The published K2K
bound $\sin^2 2\theta_{13}<0.26$ at 90\% C.L. \cite{Yamamoto:2006ty} does not translate
in any useful bound on the NSI parameter space, as we have checked
both analytically and numerically.


\section{Summary}
\label{summary}

We have presented the results of a combined analysis of the data
from \atm\ neutrinos at SuperKamiokande,  K2K and  MINOS,
performed to test neutrino nonstandard interactions. The focus
has been on the effect of adding the recent MINOS data to the
analysis of the \atm\ and K2K data.

We find that the allowed region in the space of the parameters
$\sin^2 \theta,\Delta m^2,\epee,\eptt,\epet$ has the shape
predicted by the analytics (and confirmed by the data analysis
before the MINOS data \cite{Friedland:2005vy}): it has a parabolic
direction in the space of the NSI couplings and extends to smaller
$\theta$ and larger $\dm$ with respect to the region found without
NSI. With the addition of MINOS, the region becomes narrower but
the part of the region at small $\theta$ due to NSI remains
allowed,
see Fig.~\ref{greenfigure}.

Another effect of MINOS is to shift the best fit point to non-zero NSI, mixing smaller than maximal, and slightly larger $\dm$.
However, there is only an insignificant  difference in $\chi^2$ between the best fit point and the point that minimizes the $\chi^2$ for zero NSI.

Much stronger conclusions could be obtained in the future, when
MINOS has higher statistics. Here we have shown examples of fits
to simulated MINOS data with $\simeq 27$ times larger statistics
(``MINOS-high" sample) combined with the atmospheric and K2K data as
they are today (no updates are expected from K2K, which is now
closed). We find that in this scenario the precision is sufficient
to reveal oscillation parameters in tension with the analysis of
atmospheric neutrinos in absence of NSI, and thus to give an
indication of the existence of new interactions.  It is possible
that the point $\epee=\epet=\eptt=0$ will be outside the 99\% C.L.
region of the combined analysis of all the data.  This requires a
significant shift
 of the vacuum parameters with respect to the matter ones, $\theta_m$ and $\Delta m^2_m$, corresponding to a large ``matter"
 angle, $\beta$,  $\beta \gta 0.3 \pi$ (see Eq. (\ref{cecicurve})).
 Such condition
typically corresponds to $|\epet|\gta 0.5 $ per electron, i.e. NSI
per quark at the level of $\sim 10\%$ of the Standard Model
interaction.

It should be considered that, since the sensitivity to NSI arises from
combining  data  from a neutrino beam like MINOS and the signal
 from atmospheric neutrinos, the latter  may
become the limiting factor as the precision of the beam measurements
increases.  This should motivate a second phase in the study of
atmospheric neutrino oscillations, characterized by higher precision,
desirably of the same level -- $\sim 10- 20\%$ -- of the perspective
MINOS precision.  Such second phase could be realized at Megaton water
Cherenkov detectors like UNO \cite{Jung:1999jq,Wilkes:2005rg},
HyperKamiokande \cite{Nakamura:2003hk} and MEMPHYS
\cite{Mosca:2005mi}.  Alternatively, the presence of NSI could be
revealed by the combination of high precision measurements from
different beam experiments.  A typical scenario would require two
beams of different baseline, one of similar length as MINOS and the
other with a baseline longer by a factor of several.  For this second
beam the matter width would be sufficient to have significant
refraction effects, if the epsilons are close to $\sim 1$ (per
electron, meaning about 0.2 per $u$ and $d$ quarks).  This long
baseline experiment would play the same role played by atmospheric
neutrinos in this paper.  Finally, a third possibility to test for NSI
is to use a MINOS-like beam to look at oscillation channels where
vacuum oscillations would be suppressed by the smallness of the mixing
$\theta_{13}$ rather than by the high energy, so that small effects
of NSI would not be hidden by larger vacuum oscillation effects as it
happens for part of the \atm\ \nn\ spectrum.

We have investigated the latter possibility for the case of
MINOS-high. The main finding is that epsilons of order unity could
give a measurable $\numu - \nue$ conversion probability of the
order of few $10^{-2}$.  In the interval of energy relevant for
MINOS the NSI-induced signal for $\theta_{13}=0$ can mimic the
effect of $\theta_{13}$ (with no NSI), and vice versa. To break
the degeneracy between the two scenarios would require a
combination of tests in several oscillation channels at a neutrino
beam like MINOS itself, or a quasi-vacuum test of $\numu - \nue$
conversion, of the type planned for nuclear reactor experiments,
e.g. \cite{Ardellier:2004ui}.

Finally, we again stress the importance of measuring the neutral
current event rate in the MINOS far detector, as a complementary mode
for searching for the NSI. Since the flux at the far detector
contains a significant $\nu_\tau$ component, such a measurement
would be a valuable probe of new tau neutrino interactions. This
probe would be complementary to the matter effects considered
here, because the cross-section effects are sensitive to, in
general, different combinations of the NSI as discussed in
\cite{Friedland:2005vy} \footnote{In particular, if the NSI is on
electrons, the effect on the detection cross section would be very
small.}. Such an analysis would be a generalization of the search
for the sterile neutrino component already planned by the MINOS
collaboration.

In conclusion, we have shown that the potential of MINOS goes
beyond confirming and adding precision to the already accepted
$\numu \rightarrow \nutau$ vacuum oscillations scenario that
explains atmospheric neutrinos data, and testing vacuum
oscillations in the $\numu -\nue$ channel. This experiment can be
a valuable tool to search for exotic neutrino-matter interactions,
both in the $\numu$ disappearance and in the $\nue$ appearance
parts of its program, and this should motivate its long term
operation into a phase of precision measurements.  Our study also
makes a general point, that it is important to test neutrino
oscillations with high precision both in environments where matter
effects are suppressed and in cases with strong matter effects,
because the comparison of the two could reveal new physics.  This
motivates higher precision measurements with atmospheric neutrinos
and contributes to make the case for long baseline neutrino beams
and for a new generation of reactor neutrino experiments.

\subsection*{Acknowledgments}
We are especially grateful to Michele Maltoni for  providing the
numerical executable used in our past work
\cite{Friedland:2005vy}, the results of which were used in the
present paper. A.F. was supported by the Department of Energy,
under contract number DE-AC52-06NA25396. C.L. acknowledges support
from the INT-SCiDAC grant number DE-FC02-01ER41187.

\subsection*{Note added} While the text of this work was being finalized, the
paper by Kitazawa, Sugiyama and Yasuda \cite{Kitazawa:2006iq} appeared online,
discussing ideas similar to ours about testing NSI in the $\numu
-\nue$ channel at MINOS.  Our work adds to theirs the three neutrinos
analytical formalism, helpful to analyze the $\theta_{13} - \epet$
degeneracy.

\end{document}